\documentclass[preprints,article,accept,pdftex,moreauthors]{Definitions/mdpi} 
\firstpage{1} 
\pubvolume{1}
\issuenum{1}
\articlenumber{0}
\pubyear{2022}
\copyrightyear{2022}
\datereceived{} 
\dateaccepted{} 
\datepublished{} 
\hreflink{https://doi.org/} 

\Title{Prospects of Searching for Type Ia Supernovae with 2.5-m Wide Field Survey Telescope} 

\TitleCitation{Prospects of Searching for Type Ia Supernovae with 2.5-m Wide Field Survey Telescope}

\Author{Maokai Hu $^{1,2,*}$\orcidM{}, Lei Hu $^{1,2}$\orcidL{}, Ji-an Jiang $^{3,*}$\orcidJ{}, Lin Xiao $^{4,5,6}$\orcidN{}, Lulu Fan $^{2,6}$\orcidF{}, Junjie Wei $^{1,2}$\orcidW{}, Xuefeng Wu $^{1,2}\orcidX{}$} 

\AuthorNames{Maokai Hu, Lei Hu, Ji-an Jiang, Lin Xiao, Lulu Fan, Junjie Wei, Xuefeng Wu} 

\AuthorCitation{Hu, M.; Hu, L.; Jiang, J. et al. }

\address{%
$^{1}$ \quad Purple Mountain Observatory, Chinese Academy of Sciences, Nanjing 210023, China\\
$^{2}$ \quad School of Astronomy and Space Sciences, University of Science and Technology of China, Hefei 230026, China\\ 
$^{3}$ \quad National Astronomical Observatory of Japan, National Institutes of Natural Sciences, 2-21-1 Osawa, Mitaka, Tokyo 181-8588, Japan \\
$^{4}$ \quad  Department of Physics, College of Physical Sciences and Technology, Hebei University, Wusidong Road 180, 071002, Baoding, China \\
$^{5}$ \quad Key Laboratory of High-precision Computation and Application of Quantum Field Theory of Hebei Province, Hebei University, Wusidong Road 180, 071002, Baoding, China \\
$^{6}$ \quad Department of Astronomy, University of Science and Technology of China, Hefei, 230026, China 
}   

\corres{Correspondence: kaihukaihu123@pmo.ac.cn (M. H.); jian.jiang@nao.ac.jp (J. J.)}

\abstract{Type Ia Supernovae (SNe~Ia) are the thermonuclear explosion of a carbon-oxygen white dwarf (WD) and are well-known as a distance indicator. However, it is still unclear how WDs increase their mass near the Chandrasekhar limit and how the thermonuclear runaway happens. The observational clues associated with these open questions, such as the photometric data within hours to days since the explosion, are scarce. Thus, an essential way is to discover SNe~Ia at specific epochs with optimal surveys. The 2.5-m Wide Field Survey Telescope (WFST) is an upcoming survey facility deployed in western China. In this paper, we assess the detecability of SNe~Ia with mock observations of WFST. Followed by the volumetric rate, we generate a spectral series of SNe~Ia based on a data-based model and introduce the line-of-sight extinction to calculate the brightness from the observer. By comparing with the detection limit of WFST, which is affected by the observing conditions, we can count the number of SNe~Ia discovered by mock WFST observations. We expect that WFST can find more than $3.0\times10^{4}$ pre-maximum SNe~Ia within one-year running. In particular, WFST could discover about 45 bright SNe~Ia, 99 early-phase SNe~Ia, or $1.1\times10^{4}$ well-observed SNe~Ia with the hypothesized Wide, Deep, or Medium mode, respectively, suggesting WFST will be an influential facility in time-domain astronomy.}


\begin{document}



\section{Introduction}   

Type Ia supernovae (SNe~Ia) are the thermonuclear explosion of a carbon-oxygen white dwarf (WD) in a close binary system. Of particular interest is how the WD acquires the mass near the Chandrasekhar limit from its donor star \cite{2014ARA&A..52..107M}. In the single degenerate scenario, the WD accretes matter from a main-sequence star or red giant through the Roche lobe or stellar wind \cite{1973ApJ...186.1007W,1982ApJ...253..798N}. In contrast, the double degenerate scenario suggests that the donor star is another WD and SNe~Ia originate from the merger of the two WDs \cite{1984ApJS...54..335I,1984ApJ...277..355W}. Cosmologically, SNe~Ia serve as a distance indicator based on the luminosity-width relationship \cite{1993ApJ...413L.105P,WangL:2003ApJ...590..944W,He2018ApJ...857..110H}. The measurements of SNe~Ia have exposed the accelerating expansion of the universe and constrained the properties of dark energy \cite{1998AJ....116.1009R,2007ApJ...659...98R,1999ApJ...517..565P}. Astrophysically, SNe~Ia can inject metal elements into interstellar environments and play an important role in galaxy evolution \cite{1986A&A...154..279M,1995MNRAS.277..945T,2006MNRAS.371.1125S}. SNe~Ia also tightly connect with stellar evolution, accreting process, and merger physics. 


To date, the physical nature of SNe~Ia, however, is still elusive, such as the pathway of WDs acquiring mass \cite{2014ARA&A..52..107M}, the region of thermonuclear ignition \cite{2006NewAR..50..470H}, and the existence of a surviving companion or the circumstellar material (CSM) \cite{2012Natur.489..533G,2022ApJ...931..110H}. One efficient way to reveal these mysteries is to capture the multi-band photometric signals of SNe~Ia soon after the explosion. Here are several examples: 
\begin{itemize}
\item The photometric signals within a few hours after the explosion strictly constrain the progenitor of SN~2011fe to be a WD \cite{2011Natur.480..344N}. 

\item The early-phase declining in the ultra-violent bands of iPTF14atg possibly relates to the ejecta-companion interaction \cite{2015Natur.521..328C}. 

\item The early light curve excess and red color evolution of MUSSES~1604D supports the helium burning on the surface of a WD \cite{2017Natur.550...80J}.

\item The early ultra-violet/optical bump of a large fraction of 91T/99aa-like luminous SNe~Ia suggests radioactive decay from abundant $^{56}$Ni at the outermost layer of ejecta \cite{2018ApJ...865..149J}.

\item The prominent optical flash within the first day after the explosion points to the presence of ejecta-CSM interaction for SN~2020hvf \cite{2021ApJ...923L...8J}.

\item The early-phase observations of SNe~2012cg, 2017cbv, 2018oh, 2019np, 2019yvq, and 2021aefx are also valuable for studying their physical origins \cite{2016ApJ...820...92M,2017ApJ...845L..11H,2019ApJ...870L...1D,2022MNRAS.514.3541S,2021ApJ...919..142B,2022ApJ...932L...2A}. 
\end{itemize}  

Decades of observational efforts, particularly by virtue of the transient surveys, have been dedicated to exploring the physical nature of SNe~Ia and its cosmological applications. The Supernova Legacy Survey discovered about 1,000 SNe~Ia from 2003 to 2008, and the redshift is up to 1.0 with the 3.6-meter Canada-France-Hawaii telescope \cite{2006A&A...447...31A}. The Equation of State Supernova Cosmology Experiment program found about 200 SNe~Ia with redshift from 0.1 to 0.8 using the 4-meter CTIO Blanco telescope during its 2002-2008 running \cite{2007ApJ...666..674M,2009AJ....137.3731F}. With the Hubble Space Telescope and the Subaru Hyper Suprime-Cam, the redshift of discovered SNe~Ia has been up to above 1.5 \cite{2007ApJ...659...98R,2014ApJ...783...28G,2014AJ....148...13R,2019PASJ...71...74Y,2020ApJ...892...25J}. On the other hand, several supernova surveys are designed to catch the nearby supernovae to investigate their physical nature. The Lick Observatory Supernova Search (LOSS) is a survey project focused on nearby galaxies in the northern sky \cite{2001ASPC..246..121F}. After about ten years of running, LOSS discovered 165 SNe~Ia with multi-band observations and constructed the explosion rate of SNe~Ia in the local universe \cite{2010ApJS..190..418G}. The Palomar Transient Factory (PTF) and its successor Zwicky Transient Facility (ZTF) adopted a 1.2-meter Schmidt telescope with a wide field of view to capture time-series signals of SNe~Ia  \cite{2009PASP..121.1395L,2009PASP..121.1334R,2019PASP..131g8001G}. For example, PTF~11kx is a SN~Ia with ejecta-CSM interaction signals discovered by PTF, and ZTF found a few SNe~Ia with early-phase multi-band observations \cite{2019ApJ...886..152Y,2020ApJ...902...48B}. 

The 2.5-m Wide Field Survey Telescope (WFST) is an upcoming time-domain facility deployed at the Lenghu site of western China \cite{2016SPIE10154E..2AL,2018AcASn..59...22S,2020SPIE11445E..4AL,2022MNRAS.513.2422L,LeiWFST}. The field-of-view of WFST is 6.55 square degrees, and the $r$-band limiting magnitude with 30-second exposure can reach 22.92 mag, making WFST one of the most powerful imaging survey facilities in the northern hemisphere. WFST can discover SNe~Ia in a large distance range, which can be used for both SN science and as cosmological distance indicators. In this paper, we present a preliminary study on the ability of WFST to discover SNe~Ia based on mock observations. In Section~\ref{Section2}, we introduce the configuration of our framework, including simulations of SNe~Ia in the universe, observing conditions of the Lenghu site, and the efficiency of WFST. Section~\ref{Section3} shows the ability of WFST to discover SNe~Ia with mock observations. The discussion and conclusion are shown is Section~\ref{Section4}.

\section{Methods}      
\label{Section2} 

There are two basics for constructing a framework to simulate SNe~Ia observed by WFST. The first one is simulating the brightness of SNe~Ia based on their physical properties. The second one is estimating the limiting magnitude of WFST by considering the influence of observing conditions of the Lenghu site. With this framework, we can determine whether WFST can discover SNe~Ia at specific phases and distances, and thus estimate the number of discovered SNe~Ia by WFST under different observations. To have a reasonable order estimation, we construct a simplified framework without losing any significant factors and describe them sequentially below.

\subsection{Simulations of SNe~Ia} 
\label{sec2.1} 

We randomly generate $10^{6}$ artificial SNe~Ia at specific redshifts by following the redshift evolution of the SN Ia rate from Frohmaier et al. (2019) \cite{2019MNRAS.486.2308F}. A data-based model generates the spectra energy distribution of SNe~Ia. Considering the time dilation, redshifting, distance, and dust extinction from host galaxies and the Milky Way, we can obtain the multi-band light curves of SNe~Ia by convolving WFST filters. Here, we briefly introduce the assumptions adopted in simulating SNe~Ia.

\textbf{The volumetric rate of SNe~Ia} The volumetric rate of SNe~Ia ($r(z)$) is a function of redshift $z$ constrained by the combination of the star-formation history and the delay-time distribution between the short star formation and subsequent SN~Ia events. In our study, the redshift of the simulated SNe~Ia is less than 1.0 based on the detection limit of WFST. 
We adopt a power-law distribution to describe the SN~Ia rate as $r(z) = r_0\times(1 + z)^{\alpha}$, where $r_0$ is the local SN~Ia rate with $r_0 = 2.27 \pm 0.19 \times10^{-5}\ \text{Mpc}^{-3}\text{Year}^{-1}$, and $\alpha = 1.70 \pm 0.21$ from Frohmaier et al. (2019) \cite{2019MNRAS.486.2308F}. Note that we only consider the volumetric rate of normal SNe~Ia, excluding the peculiar ones, such as the 91bg-like, 02es-like, Ia-CSM, and super-Chandrasekhar SNe~Ia.  

\textbf{The spectral template of SNe~Ia} SALT3 provides a data-based model to estimate the observer-frame time relating to the peak brightness $t_0$, a stretch-like parameter $x_1$, a color term $c$, and a scaling factor $x_0$ \cite{2021ApJ...923..265K}. Inversely, a set of spectra energy distribution of SNe~Ia can be generated by SALT3 with predetermined parameters $t_0$, $x_1$, $c$, and $x_0$ \cite{2021MNRAS.505.2819V}. 
In our study, the parameter $t_0$ is randomly distributed from one month before to one month after the mock observation run described in Section~\ref{sec2.2}. The underlying population of $x_1$ ($P(x_1)$) is an asymmetric Gaussian distribution as below, 
\begin{equation}
\label{eq1}
P(x_1) = \begin{cases} 
\exp[-(x_1 - \bar{x_1}^2) / 2\sigma_{-}^2],\  \text{if}\  x_1 \le \bar{x_1} \\ 
\exp[-(x_1 - \bar{x_1}^2) / 2\sigma_{+}^2],\  \text{if}\  x_1 > \bar{x_1}
\end{cases} 
\end{equation} 
where $\bar{x_1} = 0.938 \pm 0.101$, $\sigma_{-} = 1.551 \pm 0.118$, and $\sigma_{+} = 0.269 \pm 0.078$ \cite{2016ApJ...822L..35S}. 
The potential distribution of the color term $c$ ($P(c)$) should be a convolution between Gaussian and exponential functions corresponding to the intrinsic scatter of SN~Ia color and the extinction from the host galaxy, respectively. For simplicity, we adopt an asymmetric Gaussian function similar to Equation~\ref{eq1} to describe $P(c)$ with $\bar{c} = -0.062 \pm 0.016$, $\sigma_{-} = 0.032 \pm 0.011$ and $\sigma_{+} = 0.25$ \cite{2016ApJ...822L..35S,2014AJ....148...13R}, in which the value of $\sigma_{+}$ is consistent with the previous study on the dust model of SN~Ia host galaxies \cite{2009ApJS..185...32K}. $x_0$ is a simple normalization parameter associated with the $B$-band maximum absolute magnitude of SNe~Ia, which has a typical value of $-19.31$ with a standard deviation of $0.15$ \cite{2012ApJ...745...31B}. Based on the distributions above, we randomly generate 10,000 sets of these parameters to produce the corresponding spectra energy distribution spanning from $-20$ days to $+40$ days relative to the peak brightness. Note that the epoch of $-20$ days is not the explosion time of simulated SNe~Ia since the rising time is different under different parameter configurations in the SALT3 model. In our study, the explosion time is roughly estimated by fitting the early-phase light curve with a power-law function \cite{2016ApJ...820...67Z}.  

\textbf{The Milky Way extinction} The radiation from SNe~Ia goes through the extinction from both their host galaxies and the Milky Way. As discussed above, the color term $c$ has implicitly included the effect of the dust extinction from the host galaxy. Thus, no additional host-galaxy extinction is included. Supernova surveys usually avoid directions toward the galactic disk, so the Milky Way extinction is from the observational data with galactic latitudes $|b| > 10^{\circ}$ \cite{2011ApJ...737..103S}. 

\subsection{The Observing Conditions of Lenghu Site} 
\label{sec2.2} 

The Lenghu site is located in the relatively high latitudes of western China, leading to apparent changes in the nighttime and weather conditions during the four seasons. Thus, we divide the whole time window of one-year running into six runs. This time series with six runs can reasonably deal with the changes in nighttime and weather conditions. Each run has a time window of two adjacent months, with the first consisting of January and February. Thus, WFST could continually monitor the same sky area in each run.   

\textbf{Nighttime} Astronomically, nighttime is commonly defined as the sun's altitude is more than 15 degrees below the horizon. At the Lenghu site, the longest nighttime in winter is about 11.4 hours, and the shortest in summer is only about 5.2 hours. Such noticeable change in the nighttime is unfriendly to the SN survey because the observable time varies daily, and it is difficult to repeatedly monitor the same sky area. By dividing the 1-year observations into six runs, the observable time of each day is the shortest nighttime within the run. Although this simplified process underestimates the observable time of each day, it does guarantee that the telescope can visit the same sky area in each run. 

\textbf{Weather} Of the nights at the Lenghu site, there are about $70\%$ photometric time per year \cite{2021Natur.596..353D}. The probability of observable nights in winter is particularly high and relatively low in summer. For simplicity, we use a parameter $P_{\text{obs}}$ to describe the probability that a night is observable. In our study, $P_{\text{obs}}$ equals 0.8, 0.9, 0.6, 0.4, 0.6, and 0.8 from the first run (January and February) to the last run (November and December), respectively. This approximation satisfies the $70\%$ observable time per year and is consistent with the distribution of observable nights through the accumulated measurement in Lenghu site \cite{2021Natur.596..353D}.   

\textbf{Moon phase} The full-moon light seriously affects the sky brightness, especially in the optical $u$ and $g$ bands. For simplicity, we adopt the optical $r$ band to simulate the mock observations with WFST. We assume there is no influence on the $r$-band limiting magnitude during the dark night. The $r$-band limiting magnitude will be reduced by 0.2 or 0.5 mag during the gray night or bright night, respectively.

\subsection{The Efficiency of WFST} 

\textbf{The field of view of WFST} WFST has a field of view of 6.55 square degrees covered by nine CCDs. However, there are gaps between the CCDs, making the effective field of view $\sim5.95$ square degrees. 

\textbf{Limiting magnitude} WFST can reach the limiting magnitude of 22.92 in the $r$ band with a 30-second exposure. However, various factors in real observations could reduce this limiting magnitude, including moonlight pollution, atmospheric extinction, and background noise from host galaxies. These effects are not negligible, although they could be reduced by setting a large target-moon distance, observing the sky area near the zenith, and developing more efficient algorithms to eliminate host galaxy contamination using image subtraction. For each simulated observation of WFST, we generate a random number from 0.0 to 1.0 to reduce the $r$-band limiting magnitude. Such simplification can reasonably reduce the overestimation of discovering SNe~Ia by WFST. 

\subsection{The Configurations of Mock Observations} 

In this paper, we consider three hypothesized observing modes: wide, medium, and deep observations with the abbreviations 'Wide', 'Medium', and 'Deep', respectively. A summary of the three modes is shown in Table~\ref{tab1}. Within an observing run defined in Section~\ref{sec2.2}, we assume WFST continuously monitors the pre-selected sky area with the optical $r$ band regardless of the moon phase. The Wide mode aims to cover a large sky area; therefore, the cadence is three days, and there is one visit per night for a specific point. In contrast, daily cadence, two visits per night, and 90-second exposure ensure that the Deep mode can search for SNe~Ia in an extensive distance range. The Medium mode is a relatively moderate configuration to a certain degree, in which the cadence is one day, and there is one visit per night.

\begin{table}[t] 
\caption{A summary of the three observation modes simulated in this paper.\label{tab1}}
\newcolumntype{C}{>{\centering\arraybackslash}X}
\begin{tabularx}{\textwidth}{CCCCCCC}
\toprule
\textbf{Observation mode}	& \textbf{Filter}	& \textbf{Cadence}  & \textbf{Visits per night}  & \textbf{Exposure} & \textbf{Limiting magnitude}\\
\midrule
Wide	& $r$ & three days  & 1   &  30s  & 22.92 \\
Medium	& $r$ & one day     & 1   &  30s   & 22.92 \\
Deep	& $r$ & one day     & 2   &  90s  & 23.54 \\
\bottomrule
\end{tabularx}
\end{table}
\unskip

\section{Results} 
\label{Section3}   

For the convenience of comparing the counts of discovered SNe~Ia under different hypothesized observing modes, we define the discovery time $t_{\text{dis}}$ of SNe~Ia as the epoch corresponding to the second $r$-band observation and the discovery magnitude $m_{\text{dis}}$ as the $r$-band magnitude at the epoch $t_{\text{dis}}$. This definition can reduce contamination from fake sources and moving objects in real observations. On the other hand, purely discovering SN~Ia candidates is far from adequate for SN~Ia sciences. From the perspective of identifying a supernova candidate, the discovery time $t_{\text{dis}}$ is better to be earlier than the peak brightness so that the spectroscopic observation could be triggered at the epoch around the maximum light. To reveal the physical nature of SNe~Ia, bright or early-phase ones are still very scarce, while a well-observed light curve is necessary for cosmological distance measurements. Thus, we will investigate the ability to discover SNe~Ia at specific epochs with the three mock observations, as shown below. 

\subsection{Discovering Pre-maximum SNe~Ia}

Discovering pre-maximum SNe~Ia is essential for the SN~Ia science, e.g., classification based on the spectrum near the peak light, depicting the spectral evolution covering from early-phase to late-phase epochs, and estimating a stretch-like parameter with a well-observed light curve. However, the rising time of SNe~Ia is usually less than 20 days, making discovering pre-maximum SNe~Ia difficult. 

\begin{figure}[t]
\includegraphics[width=10 cm]{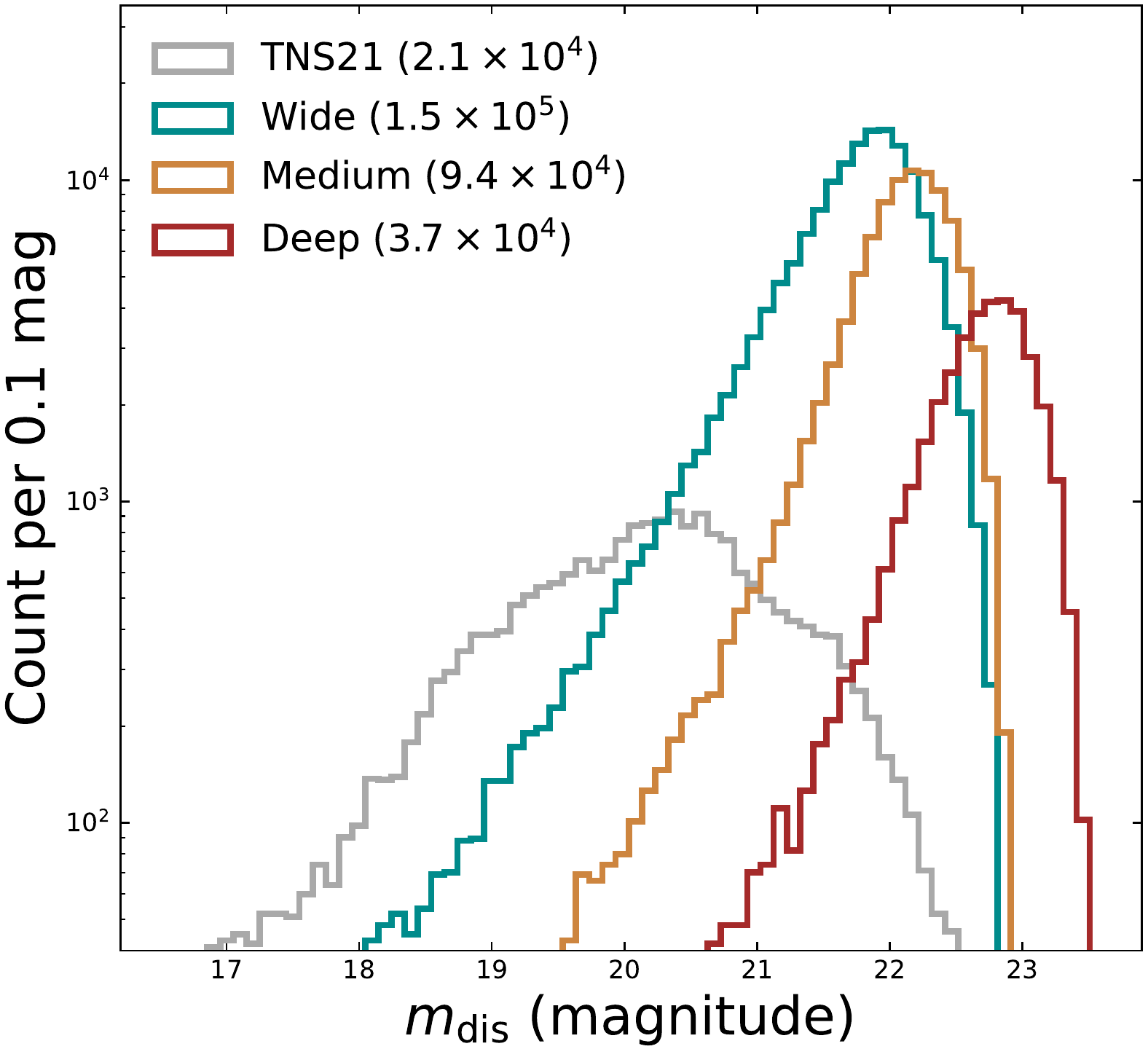} 
\caption{The gray line is the count versus the discovery magnitude ($m_{\text{dis}}$) for supernova candidates submitted to TNS in 2021 (TNS21). The teal, yellow and red lines represent the counts of pre-maximum SNe~Ia discovered by WFST under the hypothesized Wide, Medium, and Deep modes, respectively. The corresponding cumulative counts are displayed in the figure legend. \label{fig1}} \end{figure}

Transient Name Server (TNS)\footnote{https://www.wis-tns.org} is a network platform to report transients discovered by worldwide survey programs. Supernova candidates submitted to TNS in 2021 reach about $2.1\times10^{4}$, including all types of supernovae and possibly other transients, such as variables, stellar flares, novae, or active galactic nuclei. Finally, only small proportions could be classified as SNe~Ia due to the lack of spectroscopic observations. Nevertheless, the number of supernova candidates counted in TNS still provides an upper limit of the discovered supernovae. As a comparison, Figure~\ref{fig1} shows the $m_{\text{dis}}$ distributions of the discovered supernova candidates submitted to TNS in 2021 and the discovered pre-maximum SNe~Ia by WFST under the three mock observing modes. These three mock observations can discover above $3.0\times10^{4}$ pre-maximum SNe~Ia, and the cumulative count reaches $1.5\times10^{5}$ for the Wide mode. The peak of the count is at the $m_{\text{dis}}$ of 21.9, 22.2, or 22.8 magnitude for the Wide, Medium, or Deep modes, respectively. The magnitude distribution indicates that WFST can search for pre-maximum SNe~Ia in a larger distance range or discover local SNe~Ia earlier than ongoing survey programs.

With the advent of wide-field time-domain surveys, an enormous number of SNe~Ia have been discovered in the last decade. The need for extensive follow-up observations of these transients has quickly overwhelmed the limited spectroscopy resource available. Spectroscopic follow-ups will be a challenge for future WFST surveys. Utilizing machine learning classification based on photometric data can be a promising solution to fully harness the power of the WFST survey \cite{2019ApJ...884...83V}. Moreover, optimizing the spectroscopic observations for SNe~Ia may help us to efficiently utilize follow-up resources \cite{2022ApJ...930...70H}.

\subsection{Discovering Bright SNe~Ia}    

\begin{figure}[t]
\includegraphics[width=10 cm]{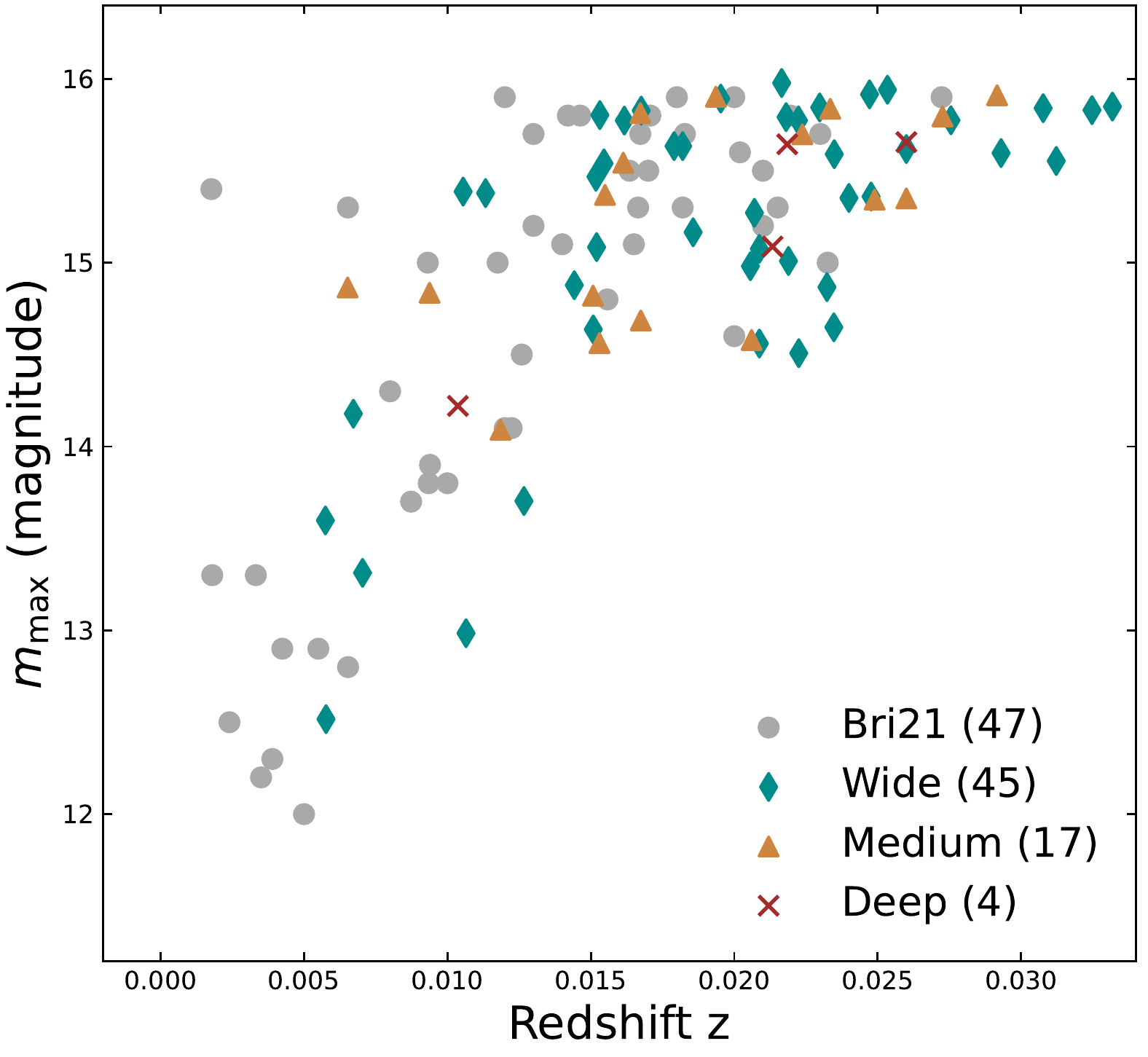}
\caption{The distributions of the maximum magnitude ($m_{\text{max}}$) versus redshift $z$ for the bright SNe~Ia defined as maximum magnitude brighter than 16.0 mag. The gray symbols are bright SNe~Ia discovered in 2021 (Bri21); the rest are the selected bright SNe~Ia discovered by WFST under the three mock observations. The total numbers of bright SNe~Ia are displayed in the figure legend.\label{fig2}} 
\end{figure}   

Bright SNe~Ia are scarce and valuable, which provide golden samples to be monitored with photometric, spectroscopic, or polarimetric observations. These diverse observations could provide rich clues to the physical origins of SNe~Ia. For instance, the asphericity derived from spectropolarimetric diagnostics strongly supports the delayed-detonation explosion of SNe~Ia \cite{2008ARA&A..46..433W,2019MNRAS.490..578C}. Furthermore, the thermonuclear ignition of carbon-oxygen WDs is off-center, suggested by the relationship between the early-phase velocity gradient of the ejecta and the late-phase velocity shift of emission lines \cite{2010Natur.466...82M}. The late-phase spectroscopic or photometric observations can also diagnose the existence of circumstellar gas around SNe~Ia, or the light echos from interstellar or circumstellar dust \cite{2007Sci...317..924P,2017ApJ...834...60Y,2022ApJ...931..110H}. 

For simplicity, we define bright SNe~Ia as the maximum brightness brighter than 16.0 magnitude. As shown in Figure~\ref{fig2}, the Wide mode could discover about 45 bright ones earlier than the peak brightness in one-year running, which is comparable to the number discovered in 2021 from the Bright Supernova website\footnote{https://www.rochesterastronomy.org/snimages/}. The total number of bright SNe~Ia is about 50 per year in the northern sky, estimated with the assumptions such as the SN~Ia rate described in Section~\ref{sec2.1}, the peak absolute magnitude of -19.3 mag, and moderate dust extinction along the line of sight. Thus, WFST and the running survey facilities can contribute to discovering bright SNe~Ia with high completeness.

\subsection{Discovering Early-phase SNe~Ia}

The early-phase SNe~Ia are full of mysteries and are regarded as a longstanding topic with theoretical or observational views. A 'dark phase' might exist between the explosion and the first light due to the adiabatic expansion of the ejecta and the absence of thermal heating from $^{56}$Ni decay. This dark phase is an essential observational phenomenon reflecting shallow or deep $^{56}$Ni deposit \cite{2013ApJ...769...67P}, which is determined by the thermonuclear runaway process of WDs. When ignition occurs at the center of a WD, a detonation will destroy the WD and produce a breakout with an X-ray flash \cite{2009ApJ...705..483H,2010ApJ...708..598P}. Besides, an ultra-violet flash might also be generated from the interaction between the ejecta and an accretion disk \cite{2009ApJ...705..483H}. However, these high-energy flashes associated with SNe~Ia have not been observed yet.   

On the other hand, several SNe~Ia events with early-phase photometric observations have shown the clues to the progenitor system or explosion mechanism, such as the helium detonation on the surface of a WD, the mixture of $^{56}$Ni to the outer ejecta, and the interaction of the ejecta with a donor star or CSM \cite{2015Natur.521..328C,2017Natur.550...80J,2021ApJ...923L...8J,2018ApJ...865..149J,2019ApJ...870L...1D}. Presently, the early-phase observations of SNe~Ia are still rare, limited by the field of view and detection limit of optical facilities and the survey mode. The advantages of WFST can make up for this shortcoming, leading to a potential opportunity for WFST to catch the early-phase signals of SNe~Ia. 

\begin{figure}[t]
\includegraphics[width=9 cm]{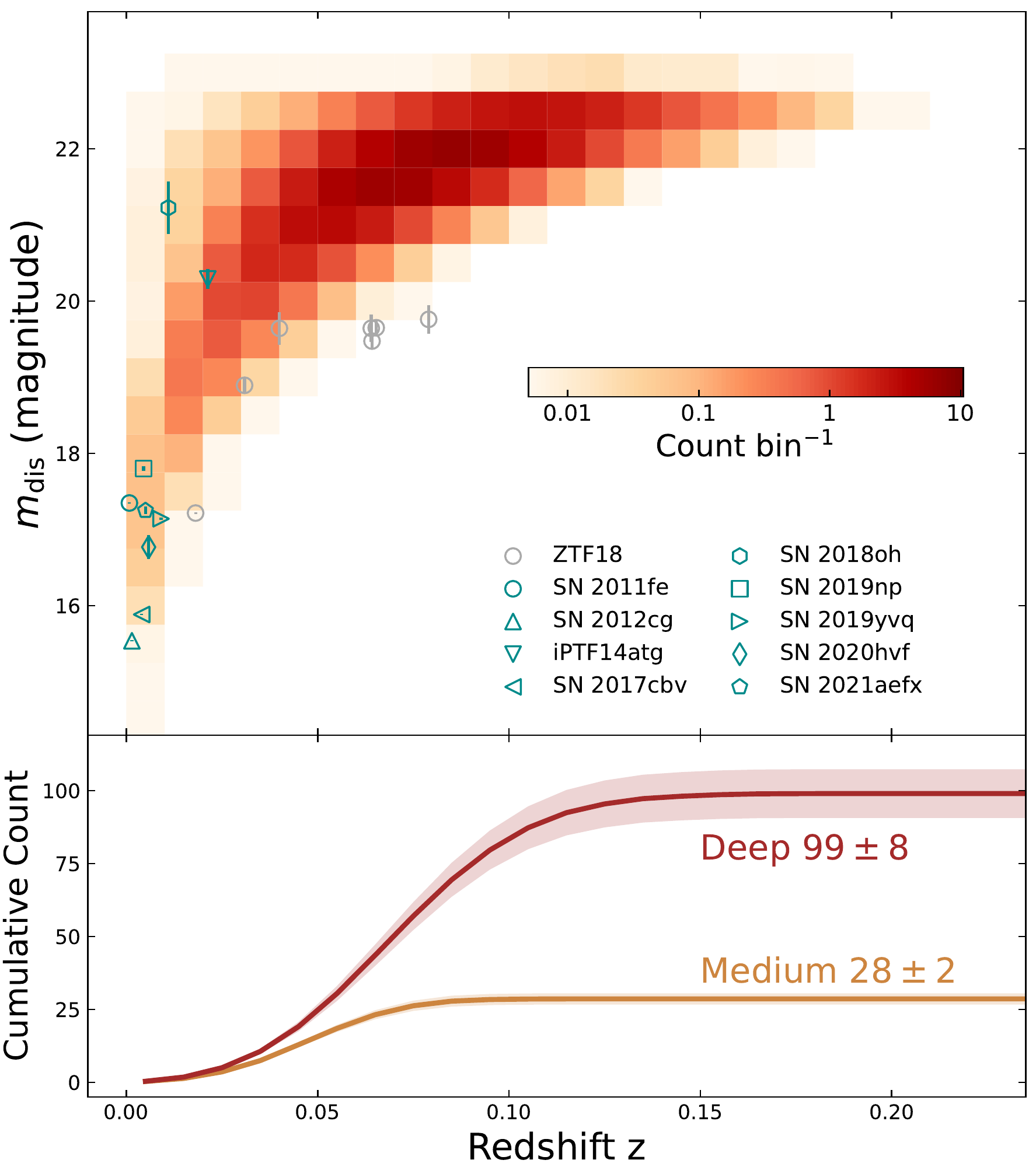}
\caption{The upper panel displays the distribution of discovery magnitude ($m_{\text{dis}}$) versus $z$ for the discovered early-phase SNe~Ia by WFST with the hypothesized Deep mode, in which each bin covers the area of 0.5 mag in $m_{\text{dis}}$ and 0.01 in $z$. The early-phase SNe~Ia is defined as $t_{\text{dis}}$ less than two days since the explosion. The gray symbols are SNe~Ia discovered by ZTF with early-phase observations, and the teal symbols are SNe~Ia with early-phase observations from other literature. The solid red line in the lower panel is the corresponding cumulative count of the discovered early-phase SNe~Ia with the hypothesized Deep mode. The yellow line represents the Medium mode for comparison. The shadow in the lower panel is the standard deviation in our simulations, which mainly corresponds to the uncertainty of the volumetric rate of SNe~Ia. \label{fig3}}
\end{figure}

We define 'early-phase SNe~Ia' as $t_{\text{dis}}$ less than two days since the explosion. Note that $t_{\text{dis}}$ is determined by the second observation, thus early-phase SNe~Ia cannot be 'discovered' by the Wide mode. The upper panel of Figure~\ref{fig3} displays the distribution of discovered early-phase SNe~Ia by the hypothesized Deep mode in the space of $m_{\text{dis}}$ versus redshift $z$. Compared with the discovered SNe~Ia with early-phase observations from ZTF or other literature \cite{2020ApJ...902...48B,2011Natur.480..344N,2015Natur.521..328C,2021ApJ...923L...8J,2016ApJ...820...92M,2017ApJ...845L..11H,2019ApJ...870L...1D,2022MNRAS.514.3541S,2021ApJ...919..142B,2022ApJ...932L...2A}, WFST can find much more early-phase SNe~Ia with higher redshift and fainter brightness. Thus, the Deep observation with WFST would contribute a substantial sample of SNe~Ia with early-phase photometric data. In comparison, the cumulative count of discovered early-phase SNe~Ia by the hypothesized Medium mode is also displayed in the lower panel of Figure~\ref{fig3}, which is less than that of the Deep mode. However, catching the signals of early-phase SNe~Ia is the first step, and the consequent efforts are also necessary, such as the rapid follow-up observations by other facilities.

\subsection{Discovering Well-observed SNe~Ia}  
\label{sec3.4} 

\begin{figure}[t]
\includegraphics[width=10 cm]{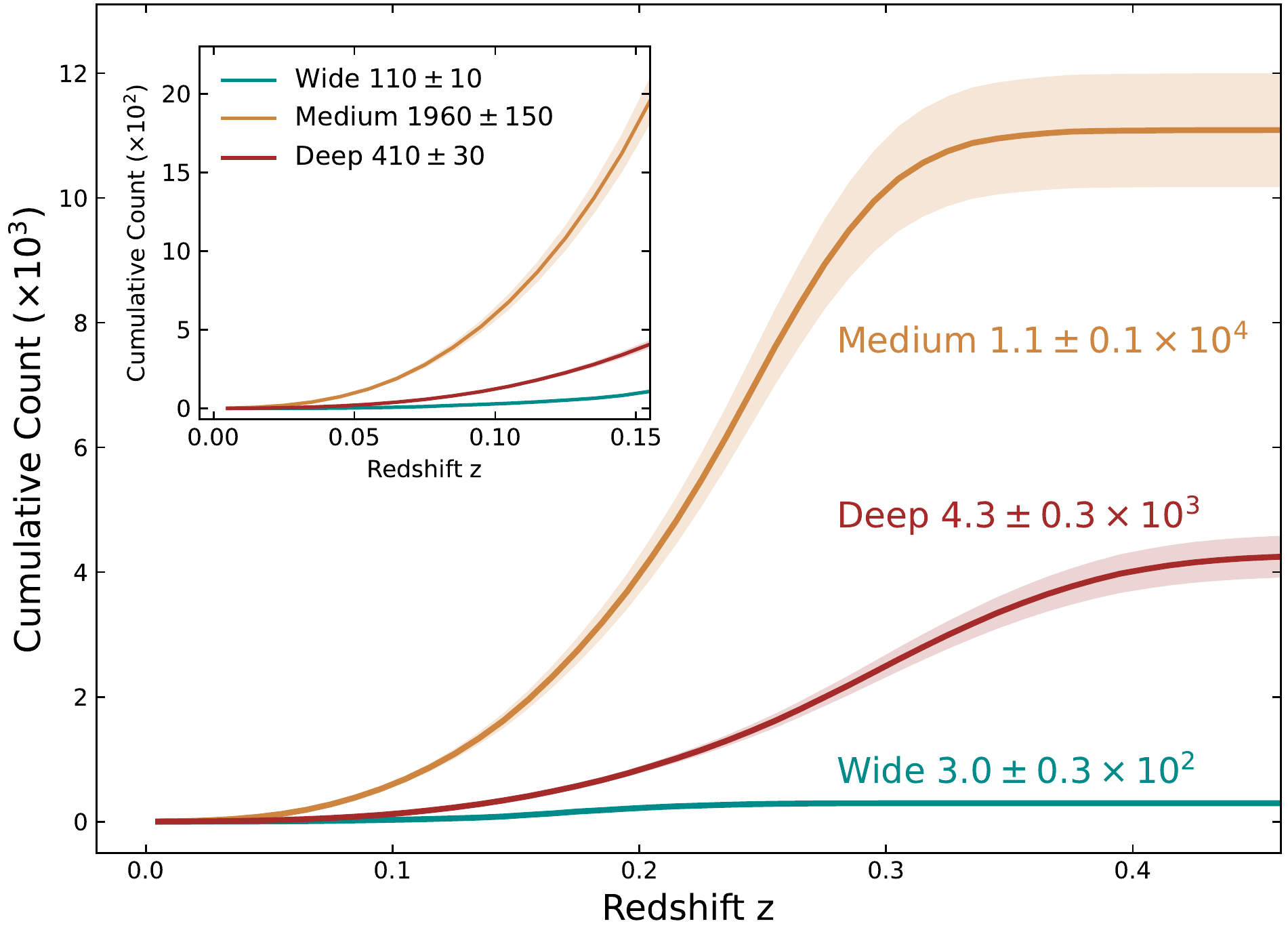}
\caption{The teal, yellow and red lines are the cumulative counts of discovered well-observed SNe~Ia with the hypothesized Wide, Medium, and Deep observations, respectively. The shadows are their corresponding standard deviation due to the uncertainty of the SN~Ia rate. The total number of well-observed SNe~Ia and the associated standard deviation are also exhibited. The insert figure highlights the cumulative count of SNe~Ia  at $z < 0.15$. \label{fig4}}
\end{figure}  

With the rapid development of transient surveys, the SNe~Ia sample has been significantly expanded, and the physical nature of SNe Ia and their cosmological applications have been widely investigated. For instance, nearby SNe Ia can be used to calibrate the Hubble constant $H_0$. The latest value of $H_0$ ($73.04\ \pm\ 1.04\ \text{km}\ \text{s}^{-1}\ \text{Mpc}^{-1}$) \cite{2009ApJ...699..539R,2011ApJ...730..119R,2018ApJ...869...56B,2018ApJ...855..136R,2019ApJ...876...85R,2022ApJ...934L...7R} measured from local SNe Ia, calibrated by the Cepheid distance ladder, is in $5\sigma$ tension with the predicted $H_0$ from the cosmic microwave background observations \cite{2014A&A...571A..16P,2020A&A...641A...6P,2021ApJ...919...16F}. For those SNe Ia at high-redshifts, they can be applied to constrain the property of dark energy \cite{2007ApJ...659...98R}, test the cosmic distance duality relation \cite{2011ApJ...729L..14L,2022ChJPh..78..297H,2022ApJ...939..115X}, constrain cosmic opacity \cite{2019ApJ...876...66W}, estimate the time variation of Newton's constant $G$ \cite{2018JCAP...10..052Z}, and so on. WFST could discover many well-observed SNe Ia with redshifts above 0.5, providing a vital opportunity for cosmological applications. On the other hand, based on the future sample of the well-observed SNe Ia from WFST, one could also expect the detection of strongly lensed SNe Ia \cite{2019ApJS..243....6G}. These events open an exciting possibility to test the Friedman-Lemaitre-Robertson-Walker metric \cite{2019PhRvD.100b3530Q}, investigate the cosmic opacity \cite{2019ApJ...887..163M}, constrain the speed of light over cosmological distances \cite{2018ApJ...867...50C}, and measure Hubble constant and cosmic curvature in a model-independent way \cite{2022PhRvD.106b3520Q}. However, only two strongly lensed SNe~Ia, iPTF16geu \cite{2017Sci...356..291G} and SN Zwicky \cite{2022arXiv221100656G}, have been discovered with multiply-imaged observations. Identifying strongly lense SNe~Ia is difficult, as these systems could not be distinguished in the usual optical transient surveys. A photometric method is to select the SNe~Ia far brighter than the normal ones as the candidates of strongly lensed SNe~Ia \cite{2017ApJ...834L...5G,2019ApJS..243....6G}. WFST could find a certain number of lensed SNe~Ia due to the detection capability. However, this part of the exciting forecasting work is beyond the scope of the simulation in this paper.  To satisfy the cosmological applications of SNe~Ia, we discuss discovering well-observed SNe~Ia by WFST in the mock observations.

The well-observed light curve is the primary condition for SNe~Ia to be used as a distance indicator because calibrating the peak luminosity of SNe~Ia is based on the shape of light curves. It is trivial to define the 'well-observed light curve' strictly, but the light curve should be better to cover the whole epochs from the rising phase to the declining phase. Thus, we set up four criteria to select the discovered SNe~Ia within the three mock observations:
\begin{itemize}
\item $t_{\text{dis}}$ is earlier than the peak brightness, which means there are at least two observations during the rising phase of SNe~Ia.  
\item there are at least two nights of observations from $-2$ days to $+2$ days relative to the peak brightness so that the maximum magnitude can be estimated properly. 
\item there are at least two nights of observations at the epoch from $+15$ days to $+30$ days after the peak brightness so that the decline of the light curve could be estimated properly.
\item there are at least fifteen nights of observations of the whole light curve so that the photometric data is sufficient to derive the light curve parameters.
\end{itemize}  

The cumulative counts of the well-observed SNe~Ia discovered by the three hypothesized observing modes are shown in Figure~\ref{fig4}. The Medium mode can find $1.1 \pm 0.1 \times 10^{4}$ well-observed SNe~Ia with the redshift up to 0.3. For cosmological use, light curves with at least three filters are necessary to calibrate the photometric magnitude to the standard optical filters in the rest-frame coordinate system and then to estimate rest-frame peak luminosity. By assuming three bands in the mock observations, the number is $650 \pm 50$ for well-observed SNe~Ia within the Hubble flow ($z < 0.15$) under the hypothesized Medium mode, which is larger than the sample used to measure $H_0$ in previous studies \cite{2022ApJ...934L...7R}. Thus, SNe~Ia search with WFST could play an important role in understanding the Hubble tension.

\section{Discussion and Conclusion} 
\label{Section4}   

In this paper, we constructed a framework to assess the ability of searching for SNe~Ia with WFST. However, several factors are not considered, which lead to a larger uncertainty of the estimated number of SNe~Ia discovered by WFST under different mock observations, as discussed below.      
\begin{itemize}
\item Although the dust extinction of host galaxies is already implied in the parameter $c$ of the SALT3 model, it is likely to underestimate the degree of host galaxy extinction. The distribution of the parameter $c$ adopted in our simulations does not correspond to highly reddened SNe~Ia. Besides, the extinction law of SN~Ia host galaxy may differ from that of the Milky Way, which makes the parameter $c$ incomplete to describe the host galaxy extinction. 
\item The explosion rate of SNe~Ia is likely correlated to the position in host galaxies and varies for different galaxy types. However, the configuration of the explosion rate in this work is only a function of redshift. The over-simplification may bring additional uncertainties into our simulations.
\item Detecting a transient close to the center of the host galaxy may involve more difficulties in data processing. The Poisson noise of the host galaxy, as well as the typical artifacts on difference images induced by inaccurate image subtraction or astronomical misalignment \cite{2022ApJ...936..157H}, can significantly degrade the true detection efficiency in a real survey.
\item Although the angular separation between the moon and the pointing of WFST is essential for attenuating moonlight contamination, the separation is ambiguous as the observed sky areas are not specified in the hypothesized modes. For simplicity, we adopt moderate values to account for the influence of the limiting magnitude by moonlight. For the same reason, the airmass is also uncertain, and its influence on the limiting magnitude is attributed to the random number ranging from 0.0 to 1.0. 
\item We simplified the influence of the weather because the weather can also affect the limiting magnitude, which is not considered in this study. 
\item The optical $r$ band is the only filter considered in this paper. For observations with more filters, the covered sky area per night should be reduced accordingly. 
\end{itemize}  

Nevertheless, our simulations demonstrate the impressive performance of WFST in searching for SNe~Ia, as WFST can find over $3.0\times10^{4}$ pre-maximum SNe~Ia within one-year running under the hypothesized Wide, Medium, or Deep modes. Specifically, the Wide mode has an advantage in discovering bright SNe~Ia with a total number of about 45 per year; it is about 99 per year for the discovered early-phase SNe~Ia under the Deep mode; the Medium mode can find about $1.1\times10^{4}$ SNe~Ia with well-observed light curves. Therefore, a concrete observing plan of WFST needs to be determined through comprehensive arguments, such as considering the scientific goals, observing conditions of the Lenghu site, and the operating state of WFST.


\vspace{6pt} 



\authorcontributions{Conceptualization, M.H. and J.J.; methodology, M.H., L.H., J.J., and L.X.; software, M.H.; validation, L.H. and J.J.; resources, L.F., J.W., and X.W.; data curation, M.H.; writing—original draft preparation, M.H.; writing—review and editing, L.H., J.J., J.W., and L.X.; visualization, M.H.; supervision, X.W.; project administration, M.H., J.J., and X.W.; funding acquisition, X.W. All authors have read and agreed to the published version of the manuscript.} 


\funding{This research was funded by the National Key Research and Development Programs of
China (2022SKA0130100), the National Natural Science Foundation of China (grant Nos. 11725314, and 12041306), the Key Research Program of Frontier Sciences (grant No. ZDBS-LY- 7014) of Chinese Academy of Sciences, the Major Science and Technology Project of Qinghai Province (2019-ZJ-A10), the Natural Science Foundation of Jiangsu Province (grant No. BK20221562), and the China Manned Space Project (CMS-CSST-2021-B11).}  

\institutionalreview{Not applicable} 

\informedconsent{Not applicable} 
 
\dataavailability{Some data used in this work are available from the published literature.} 

\acknowledgments{We gratefully thank L. Wang for the instructive comments on the background of Type Ia Supernovae and thank Q. Zhu for providing the information about WFST. Maokai Hu thanks to Jiangsu Funding Program for Excellent Postdoctoral Talent. Lei Hu acknowledges support from Jiangsu Funding Program for Excellent Postdoctoral Talent and China Postdoctoral Science Foundation (Grant No. 2022M723372). Xiao Lin is thankful for the support from National Natural Science Foundation of China (grant No. 12103050 ), Advanced Talents Incubation Program of the Hebei University, and Midwest Universities Comprehensive Strength Promotion project.}

\conflictsofinterest{The authors declare no conﬂict of interest.} 



\abbreviations{Abbreviations}{
The following abbreviations are used in this manuscript:\\

\noindent 
\begin{tabular}{@{}ll}
MDPI & Multidisciplinary Digital Publishing Institute\\
SNe~Ia & Type Ia Supernovae \\ 
WD    &  White Dwarf  \\ 
$H_0$ &  Hubble Constant \\ 
CSM &  Circumstellar Material \\ 
LOSS  &  Lick Observatory Supernova Search \\ 
PTF   &  Palomar Transient Factory \\ 
ZTF   &  Zwicky Transient Facility   \\ 
WFST  &   2.5-m Wide Field Survey Telescope \\ 

\end{tabular}
}


\begin{adjustwidth}{-\extralength}{0cm}

\reftitle{References}




\bibliography{bibtex} 

%


\end{adjustwidth}
\end{document}